\documentclass[PRD,showpacs,superscriptaddress,twocolumn]{revtex4}

\usepackage{amsmath,bm}
\usepackage{graphicx}
\usepackage{epsfig}	
\usepackage{multirow}
\usepackage{color}
\usepackage[breaklinks,colorlinks,urlcolor=blue,citecolor=blue,linkcolor=blue]{hyperref}
\usepackage{palatino}
\usepackage{verbatim}

\begin{document}

\title{Substructures of heavy flavor jets in $\bm{pp}$ and PbPb collisions at $\bm{\sqrt{s}}$ = 5.02~TeV}

\date{\today  \hspace{1ex}}
\author{Qing Zhang}
\affiliation{Key Laboratory of Quark \& Lepton Physics (MOE) and Institute of Particle Physics, Central China Normal University, Wuhan 430079, China}

\author{Zi-Xuan Xu}
\affiliation{School of Mathematics and Physics, China University of Geosciences, Wuhan 430074, China}

\author{Wei Dai}
\email{weidai@cug.edu.cn}
\affiliation{School of Mathematics and Physics, China University of Geosciences, Wuhan 430074, China}

\author{Ben-Wei Zhang}
\email{bwzhang@mail.ccnu.edu.cn}
\affiliation{Key Laboratory of Quark \& Lepton Physics (MOE) and Institute of Particle Physics, Central China Normal University, Wuhan 430079, China}

\author{Enke Wang}
\affiliation{Guangdong Provincial Key Laboratory of Nuclear Science, Institute of Quantum Matter, South China Normal University, Guangzhou 510006, China}
\affiliation{Key Laboratory of Quark \& Lepton Physics (MOE) and Institute of Particle Physics, Central China Normal University, Wuhan 430079, China}

\begin{abstract}

Groomed jet substructure measurements, the momentum splitting fraction $z_g$ and the groomed jet radius $R_g$, of inclusive, D$^0$-tagged and B$^0$-tagged jets in $pp$ and central PbPb collisions at $\sqrt{s}=5.02$~TeV are predicted and investigated. Charged jets are constrained in a relatively low transverse momentum interval 15 $\leq p_{\rm T}^{\rm jet\ ch} <$ 30 GeV/$c$ where the QCD emissions are sensitive to mass effects. 
In $pp$ collisions, heavy flavor jets have less momentum-balanced splittings than inclusive jets and the larger quark mass is, the less momentum-balanced splittings are. B$^0$-tagged jets distribute at lager $R_g$ than both D$^0$-tagged and inclusive jets, while D$^0$-tagged jets distribute at smaller $R_g$ than inclusive jets. That can be explained by the competition between parton mass effects and flavour effects on splitting-angle distributions. In $A$+$A$ collisions, jet quenching effects further suppress the momentum-balanced splittings and enhance the large-angle splittings. Heavy flavor jets suffer more $z_g$ and less $R_g$ modifications due to jet-medium interactions. The mass hierarchy in $z_g$ of inclusive, D$^0$-tagged, B$^0$-tagged jets and competition between mass effects and Casimir colour factors in $R_g$ can also be observed in PbPb collisions.
\end{abstract}


\maketitle

\section{Introduction}
\label{sec:intro}

\begin{figure*}[t]

    \begin{minipage}{1\linewidth}
		\begin{center}
			\resizebox{0.99\textwidth}{!}{
				\includegraphics{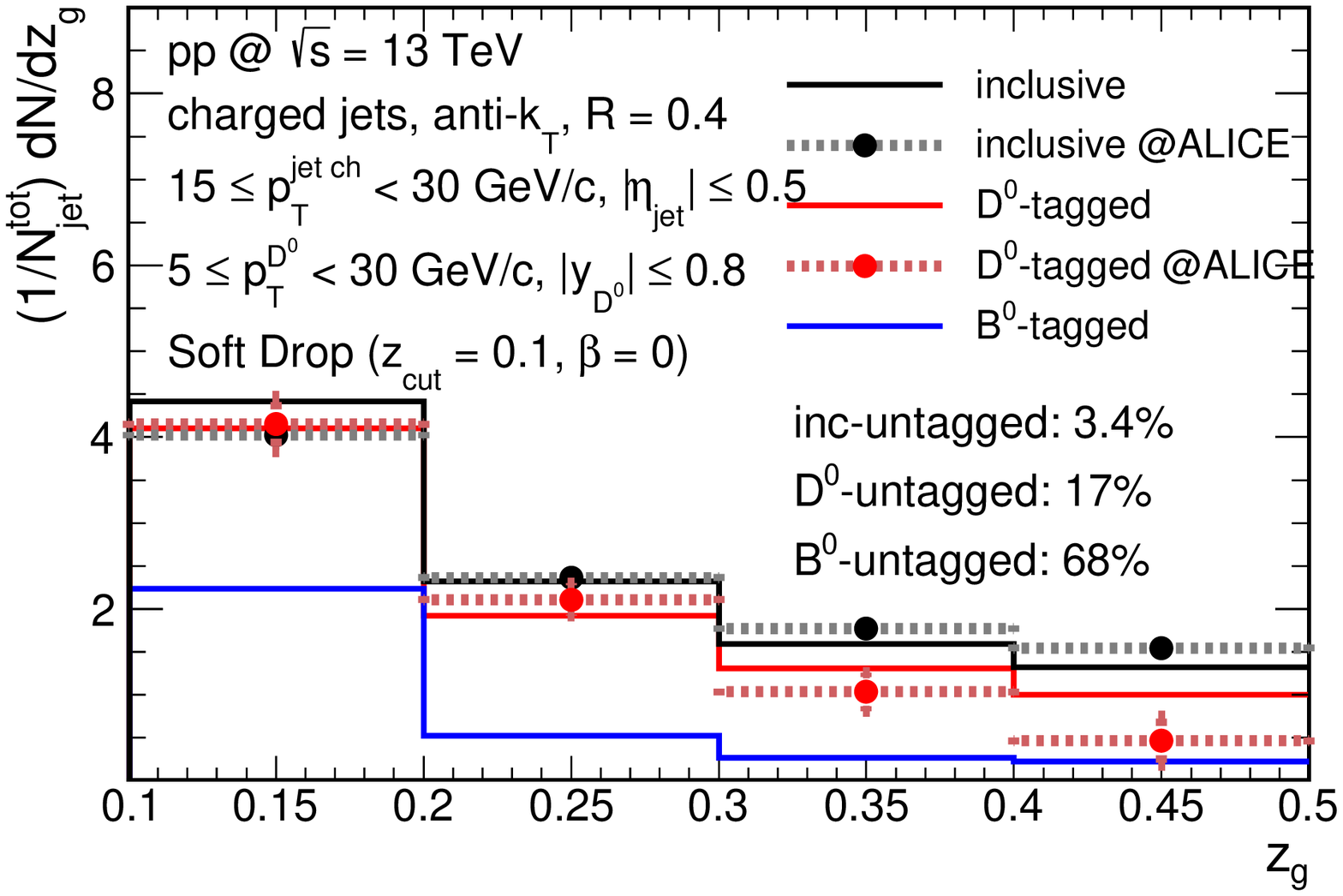}
				\includegraphics{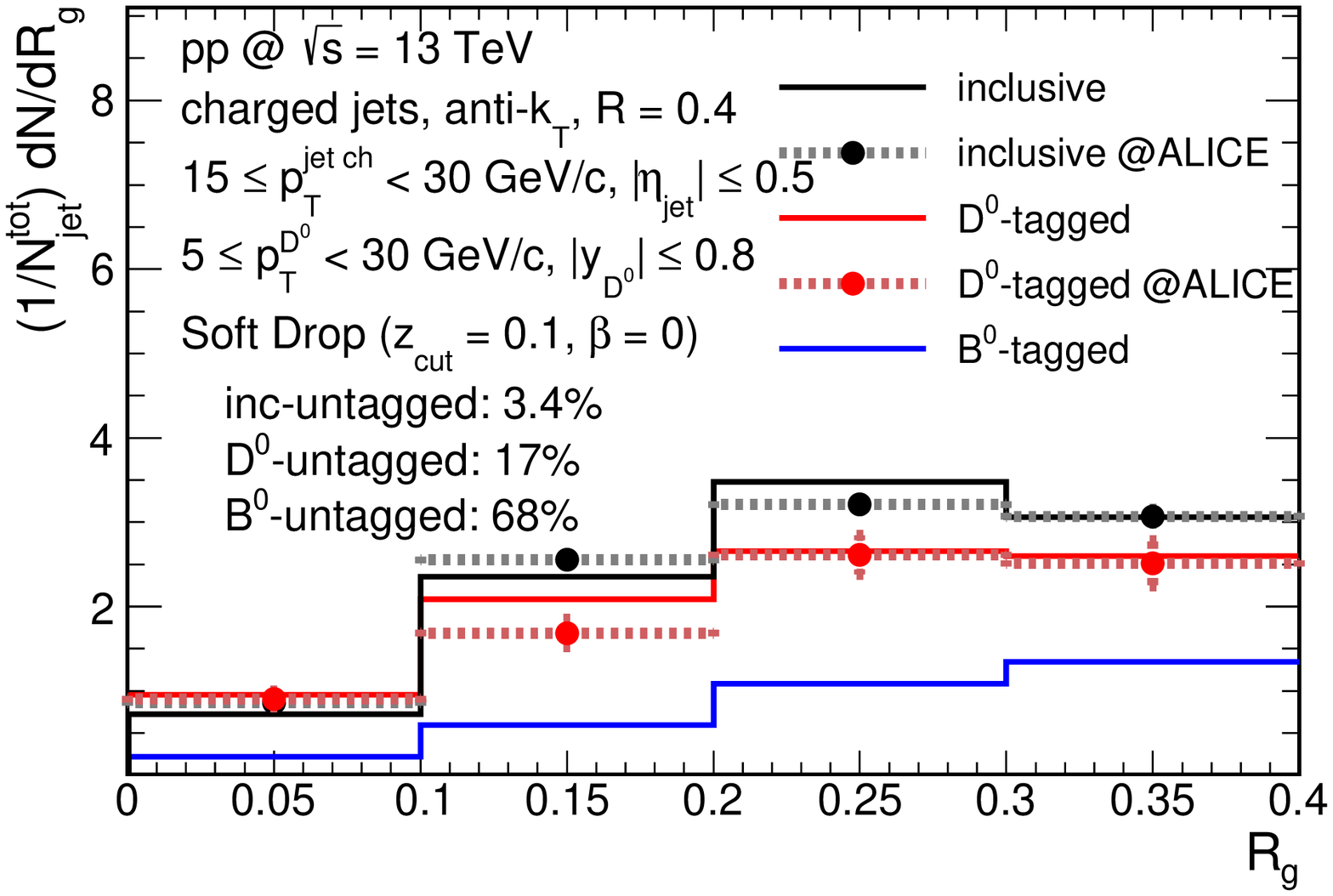}
			}
		\end{center}
	\end{minipage}
	
	\caption{(color online) $z_g$ (left) and $R_g$ (right) distributions for inclusive and D$^0$-tagged jets in 15 $\leq p_{\rm T}^{\rm jet\ ch} <$ 30 GeV/$c$ in $pp$ collisions at $\sqrt{s}$ = 13~TeV, normalised to the total number of jets $N_{\rm jet}^{\rm tot}$, are compared to ALICE data (dotted line)~\cite{ALICE:2022phr}. Predictions of distributions for B$^0$-tagged jets are also plotted.}
	\label{fig:pp13}
\end{figure*}

The quark-gluon plasma (QGP) created in the ultra-relativistic heavy-ion collisions (HIC) has inspired great interest in the last two decades~\cite{Shuryak:2014zxa, Braun-Munzinger:2015hba, Busza:2018rrf}. Jets, the collimated sprays of particles initiated by high-energy quarks and gluons may lose energy when traversing through the QGP medium which has been referred to as the jet quenching effects~\cite{Wang:1992qdg, Zhang:2003wk, Gyulassy:2003mc, Qin:2015srf, Ma:2010dv, Fochler:2011en}. The impact of the different Casimir color factors of quark and gluon, and the impact of the quark masses on the jet quenching effect, are essential to understand the full nature of jet-medium interaction. It is expected that gluons will lose more energy than light quarks during the medium-induced interactions due to different Casimir color factors, and the massive quarks will lose less energy than light quarks due to the fact that energies carried away by the gluon radiations from massive quarks suffering a suppression in small-angle determined by their finite masses~\cite{Buzzatti:2011vt, Andronic:2015wma}, which is called as the dead-cone effect. The recent development of the jet substructure measurements provide fine tool to conduct such investigations since they are closer to the initial hard process~\cite{Larkoski:2014wba}. 

The soft-drop groomed jet substructure measurements, the groomed shared momentum fraction $z_g$, and the groomed jet radius $R_g$, have been proven successful in characterizing the hard two-prong substructure of charged jets. They had been calculated and experimentally measured both in $pp$ and in PbPb~\cite{STAR:2020ejj, CMS:2018ypj, ATLAS:2019mgf, ALICE:2021njq, ALICE:2022phr, ALICE:2022hyz, CMS:2014jjt, CMS:2017qlm, STAR:2021kjt, ATLAS:2017nre, ALICE:2019ykw, ALargeIonColliderExperiment:2021mqf, ALICE:2022vsz, Wang:2022yrp}. It is natural to investigate the case in heavy flavor jets. In a recent experiment measurement~\cite{ALICE:2021aqk}, in which direct observation of the Dead-Cone effect in a vacuum has been reported, relatively low transverse momentum intervals of the charm-meson tagged jets are constrained where the mass effect is mostly pronounced.  
Inspired by this, $z_g$ and $R_g$ distributions of $D^0$-tagged charged jets are measured by ALICE in $pp$ collisions at $\rm \sqrt{s_{NN}}=13$~TeV with the requirement of $p_{\rm T}^{\rm jet\ ch}<$~30~GeV/$c$~\cite{ALICE:2022phr}. In the same dynamic region, the case in the inclusive charged jets which are referred to as the no mass limit and believed to be dominated by the gluon-initiated jets is also experimentally studied. It highlights the need to further predict these soft-drop groomed jets substructure measurements of both heavy flavor jets and inclusive jets in $A$+$A$ collisions in such relatively low transverse momentum intervals of jets. Interestingly, such prediction will help isolate the mass and flavor dependent properties of the QCD emissions in the same dynamic region.

In order to do so, in this manuscript, the groomed jets substructure of Beauty-meson tagged charged jets will also be predicted both in $pp$ and $A$+$A$ to explore the parton mass hierarchy, and a systematic comparison with charm-meson tagged and inclusive jets will be conducted. Therefore, the rest of this paper is organized as follows: in Sec.~\ref{sec:pp} we first calculate the groomed jet substructure observables, $z_g$ and $R_g$ of $D^0$-meson tagged jets and inclusive charged jets constrained by 15 $\leq p_{\rm T}^{\rm jet\ ch} <$ 30~GeV/$c$ in $pp$ collisions at $\sqrt{s}$ = 13~TeV with PYTHIA~8 Monte Carlo event generator confronted with ALICE data. Then predictions of those observables for $B^0$-meson tagged jets in $pp$ are provided in the same kinematics range. In Sec.~\ref{sec:AA} the theoretical framework of SHELL model is utilized to describe the in-medium evolutions of massive and massless parton simultaneously. We calculate $z_g$ and $R_g$ distributions for inclusive, $D^0$-meson tagged and $B^0$-meson tagged jets in PbPb collisions at $\sqrt{s}$ = 5.02~TeV in the same kinematics region. And the mass and flavour dependence of medium modification are also discussed in this section. Finally we summarize and conclude in Sec.~\ref{sec:sum}.

\section{Jet substructure in $\textit{pp}$ collisions}
\label{sec:pp}

\begin{figure*}[t]

		\begin{center}
				\includegraphics[width=0.49\textwidth]{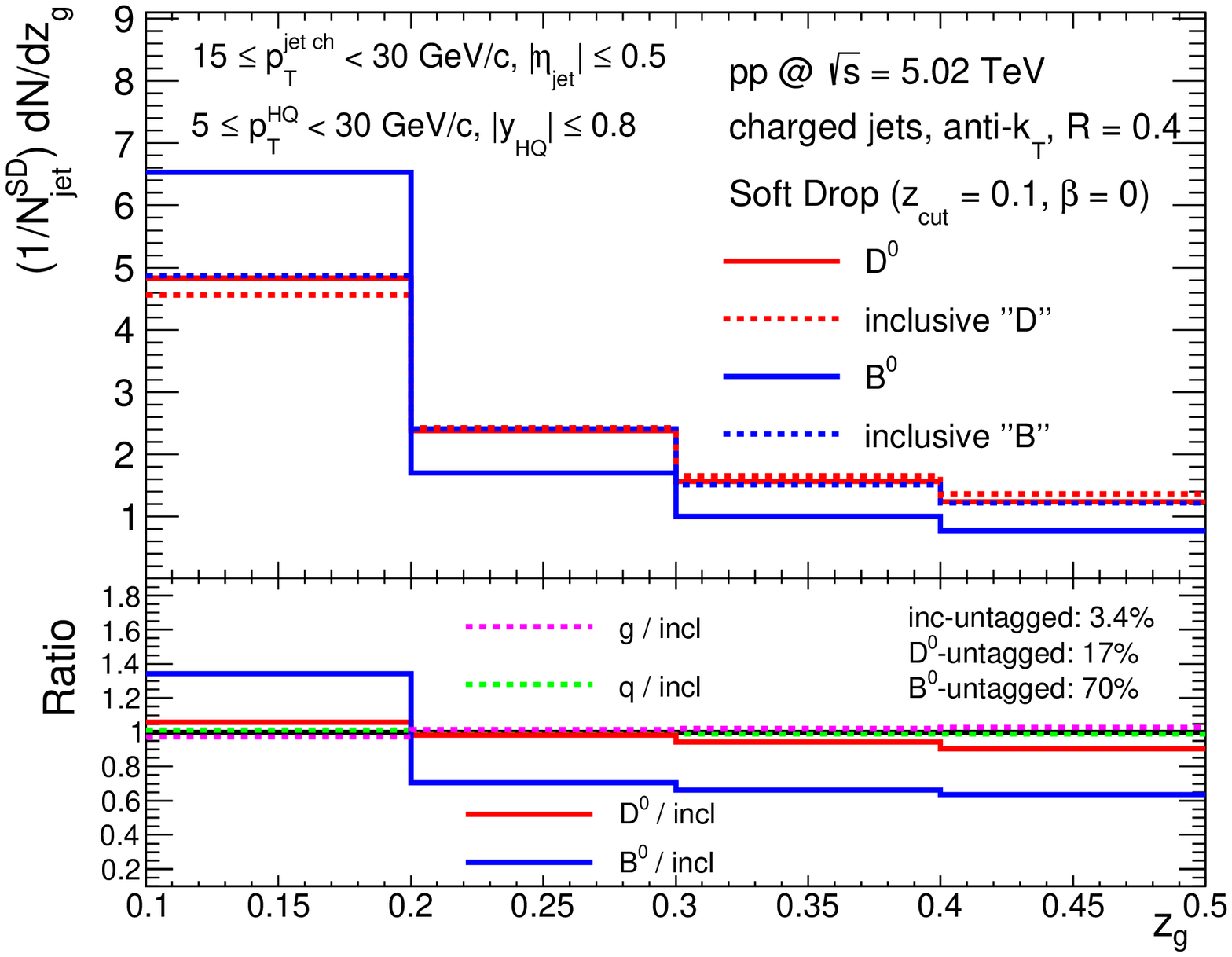}
				\includegraphics[width=0.49\textwidth]{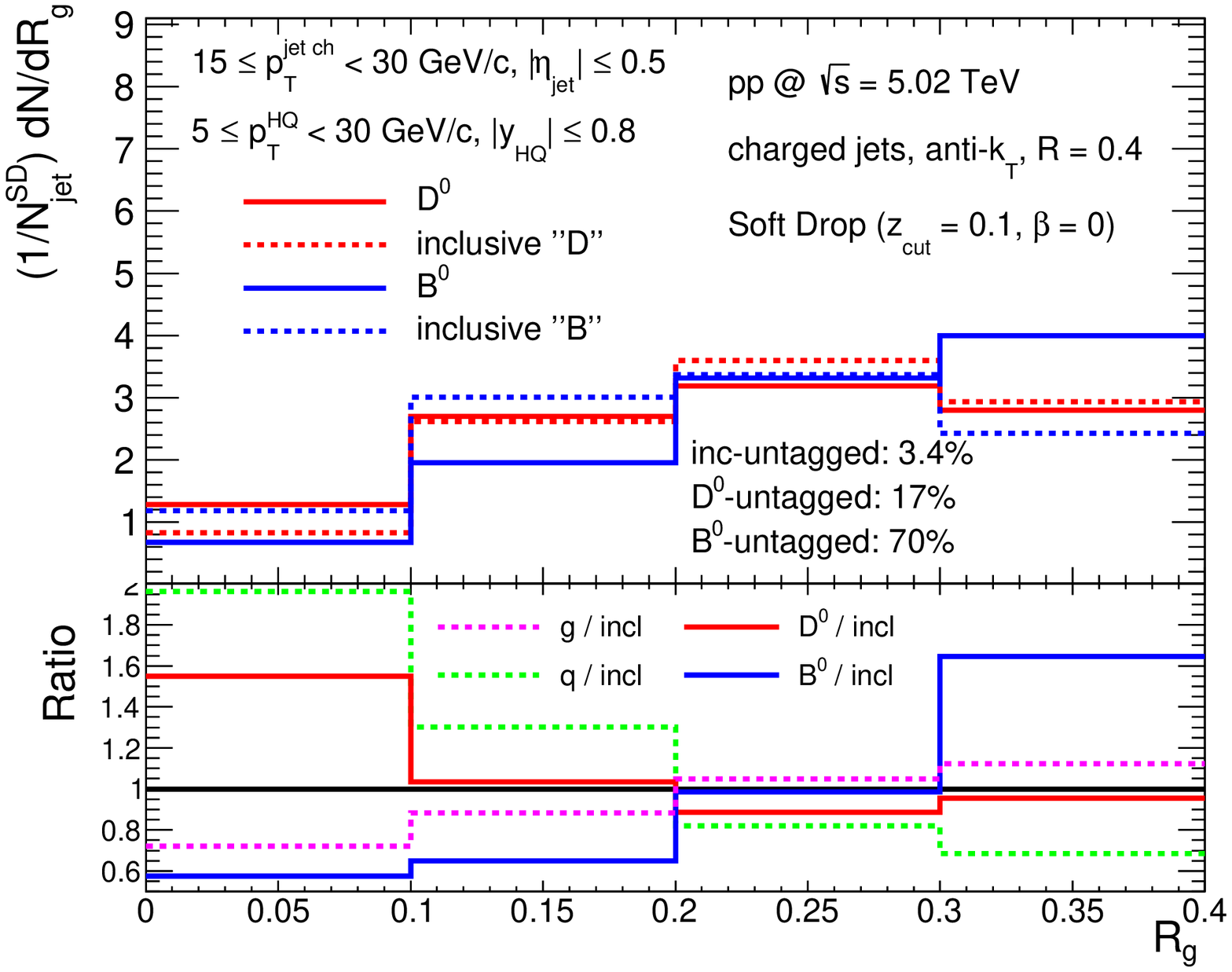}
		\end{center}
	
	\caption{(color online) Upper: Predictions of $z_g$ (left) and $R_g$ (right) distributions for D$^0$-tagged, B$^0$-tagged jets and inclusive "D", inclusive "B" jets in 15 $\leq p_{\rm T}^{\rm jet\ ch} <$ 30 GeV/$c$ in $pp$ collisions at $\sqrt{s}$ = 5.02~TeV, normalised to the number of groomed jets $N_{\rm jet}^{\rm SD}$. Bottom: $z_g$ (left) and $R_g$ (right) distribution-ratios of pure light-quark initiated, pure gluon initiated, D$^0$-tagged and B$^0$-tagged jets over those of inclusive jets.}
	\label{fig:pp4}
\end{figure*}

With the reclustering techniques, we are able to tag heavy-quark initiated jets with a heavy flavor meson. The charm-quark jets are tagged through a reconstructed D$^0$ meson, which has a mass of 1.86 GeV/$c^2$~\cite{ParticleDataGroup:2020ssz}, amongst their constituents. And a B$^0$ meson, with a mass of 5.28 GeV/$c^2$~\cite{ParticleDataGroup:2020ssz}, is requested in the constituent-lists when constraining the B$^0$-tagged jets. The same configuration as ALICE report~\cite{ALICE:2022phr} is applied in this work: The $pp$ events are generated by the Monte Carlo (MC) event generator PYTHIA8~\cite{Sjostrand:2014zea} Monash 2013 Tune~\cite{Skands:2014pea}. Then the jets are reconstructed with FastJet 3.2.1~\cite{Cacciari:2011ma} anti-$k_{\rm T}$~\cite{Cacciari:2008gp} algorithm for resolution parameter $R$ = 0.4. To manifest the parton-mass influences on the emission properties, a relatively low transverse-momentum interval 15 $\leq p_{\rm T}^{\rm jet\ ch} <$ 30~GeV/$c$ is applied. And jet pseudorapidity range is $|\eta_{\rm jet}| < 0.5$. The soft drop condition is chosen as $z_{cut}$ = 0.1 and $\beta$ = 0 in the jet grooming procedure. The tagging HF meson D$^0$ and B$^0$ (denoted as ``HQ") are selected in the transverse-momentum interval 5 $\leq p_{\rm T}^{\rm HQ} <$ 30~GeV/$c$ and rapidity range $|y_{\rm HQ}| < 0.8$. In order to make direct comparisons with D$^0$-tagged jets and isolate the mass effect, we introduce an inclusive jet reference denoted as "the inclusive 'D'", which represents the production of {\it the inclusive jet} by imposing a constraint with the transverse momentum of the leading hadron in each hardest prong  $p_{\rm T,inclusive\ jets}^{\rm ch,leading\ track}\geq$ 5.33~GeV/$c$ (equal to the transverse mass of a D$^0$ meson with transverse momentum of 5~GeV/$c$). In the same manner, we also introduce the inclusive jet reference denoted as "the inclusive 'B'", which stands for the production of the inclusive jet with the requirement $p_{\rm T, inclusive\ jets}^{\rm ch, leading\ track}\geq$ 7.27~GeV/$c$ (equal to the transverse mass of a 5~GeV/$c$ B$^0$ meson).

We calculate the $z_g$ and $R_g$ distributions of the D$^0$-tagged jets and their inclusive jets reference "$D$" in $pp$ collisions at $\sqrt{s}$ = 13~TeV to confront with ALICE data~\cite{ALICE:2022phr} in Fig.~\ref{fig:pp13}, they're normalized to the total number of jets $N_{\rm jet}^{\rm tot}$ in 15 $\leq p_{\rm T}^{\rm jet\ ch} <$ 30~GeV/$c$. $z_g$ distributions of both D$^0$-tagged and inclusive jets tend to decrease with increasing $z_g$ and the curve of D$^0$-tagged jets is steeper than inclusive jets, which indicates that the shared momentum of D$^0$-tagged jets is more imbalanced than its inclusive jets reference. D$^0$-tagged jets are comparable with their inclusive reference at small $R_g$ and lower than the inclusive case at large $R_g$ illustrating that the splitting angles of D$^0$-tagged jets are smaller than their inclusive counterparts. Then we predict the $z_g$ and $R_g$ distributions of B$^0$-tagged jets in $pp$ with the same conditions as is shown in Fig.~\ref{fig:pp13} with the blue solid line. We find that both the $z_g$ and $R_g$ distributions for B$^0$-tagged jets are much lower than those for D$ ^0$-tagged jets and the inclusive references. It is mainly due to the fact that a large fraction of B$^0$-tagged jets in 15 $\leq p_{\rm T}^{\rm jet\ ch} <$ 30~GeV/$c$ can not pass the soft drop condition. We systematically calculate the fraction of jets that do not pass the soft drop condition (named as untagged) for inclusive, D$^0$-tagged and B$^0$-tagged jets respectively, we find their untagged rates are 3.4\%, 16.6\%, and 67.5\%. Therefore in the following investigation, we uniformly normalize the $z_g$ and $R_g$ distributions to the number of soft-drop groomed jets $N_{\rm jet}^{\rm SD}$.

To provide $pp$ baseline, the $z_g$ and $R_g$ distributions for D$^0$-tagged and B$^0$-tagged jets in the same transverse momentum interval 15 $\leq p_{\rm T}^{\rm jet\ ch} <$ 30 GeV/$c$, normalized to $N_{\rm jet}^{\rm SD}$, in $pp$ collisions at $\sqrt{s}$ = 5.02~TeV are shown in the left upper panel of Fig.~\ref{fig:pp4}. The inclusive reference "D" and "B" are also demonstrated. The $z_g$ distributions of D$^0$-tagged, inclusive  "D" and inclusive "B" are very close to each other while the normalized $z_g$ distribution for B$^0$ meson-tagged jets are more imbalanced simply induced by the larger quark mass of Beauty. Ratios of D$^0$-tagged, B$^0$-tagged jets over their inclusive counterparts are then plotted in the bottom panels. One can find that the distribution for heavy-flavour jets is enhanced at small $z_g$ and suppressed at large $z_g$ compared to that for inclusive jets, and clearly, the larger quark mass is, the shared momentum-imbalance will be for those jets. The mass hierarchy manifests itself in the ordering of inclusive, D$^0$-tagged, and B$^0$-tagged jets. To facilitate further discussion, the $z_g$ distributions for pure light-quark initiated and pure gluon-initiated jets are introduced and the ratio of pure light-quark initiated over the inclusive case and also the ratio of pure gluon initiated over the inclusive case are also plotted in the left bottom panel of Fig.~\ref{fig:pp4}, those ratios are all equal to unity within uncertainties. In this transverse momentum interval 15 $\leq p_{\rm T}^{\rm jet\ ch} <$ 30 GeV/$c$, the observable $z_g$ can not serve as a tool to separate the flavour dependence of splitting structure of the jets.

Let us turn to $R_g$ plotted in the right panel. We first compare the $R_g$ distributions for D$^0$-tagged and B$^0$-tagged jets, one can find the $R_g$ distribution for B$^0$-tagged jets is lower than that for D$^0$-tagged jets at small $R_g$ and higher than that at large $R_g$. It indicates the mass effect will lead the splitting angles of those tagged jets to distribute at larger $R_g$. When we compare the $R_g$ distribution for D$^0$-tagged jets and the inclusive jets reference "D" which served as the no mass limit, the case is the opposite. The $R_g$ distribution for D$^0$-tagged jets is higher than that for inclusive jets reference "D" at small $R_g$ and slightly lower at large $R_g$. It means that the mass effect can not help the D$^0$-tagged jets to be wider than the inclusive jets. When we systematically compare B$^0$-tagged, D$^0$-tagged and their inclusive references, the mass hierarchy is no longer there. It's because of the complexity of inclusive jets, so we again introduce pure light-quark initiated and pure gluon-initiated jets for better demonstration and further analysis. Therefore we plot distribution ratios of B$^0$-tagged, D$^0$-tagged, light-quark initiated and gluon-initiated jets over their inclusive jet references simultaneously in the bottom panel, denoted as B$^0$/inclusive, D$^0$/inclusive, quark/inclusive and gluon/inclusive respectively. Firstly we compare gluon/inclusive and quark/inclusive, the gluon/inclusive is closer to unity and quark-initiated jets are narrower than the inclusive ones. It also implies that inclusive jets are mainly comprised of gluon-initiated jets, and the gluon-initiated jets have wider splitting angle than quark-initiated jets. Then we compare the three curves representing quark/inclusive, D$^0$/inclusive and B$^0$/inclusive, the mass hierarchy is re-emerging. If we simply replace the quark/inclusive with gluon/inclusive, the mass hierarchy is therefore challenged by the Casimir colour factor changed from quark to gluon. We can use the case of light quark-initiated jets as a baseline.  D$^0$ tagged jets case is quark-initiated jets plus quark mass and inclusive jets case are quark-initiated jets plus Casimir colour change, these two effects will all lead the splitting angle to be wider. So when comparing the splitting-angle distribution for inclusive jets and  D$^0$ tagged jets in the upper panel, it is these two effects that compete with each other. 

\begin{figure*}[t]

	\begin{minipage}{1\linewidth}
		\begin{center}
				\includegraphics[width=0.49\textwidth]{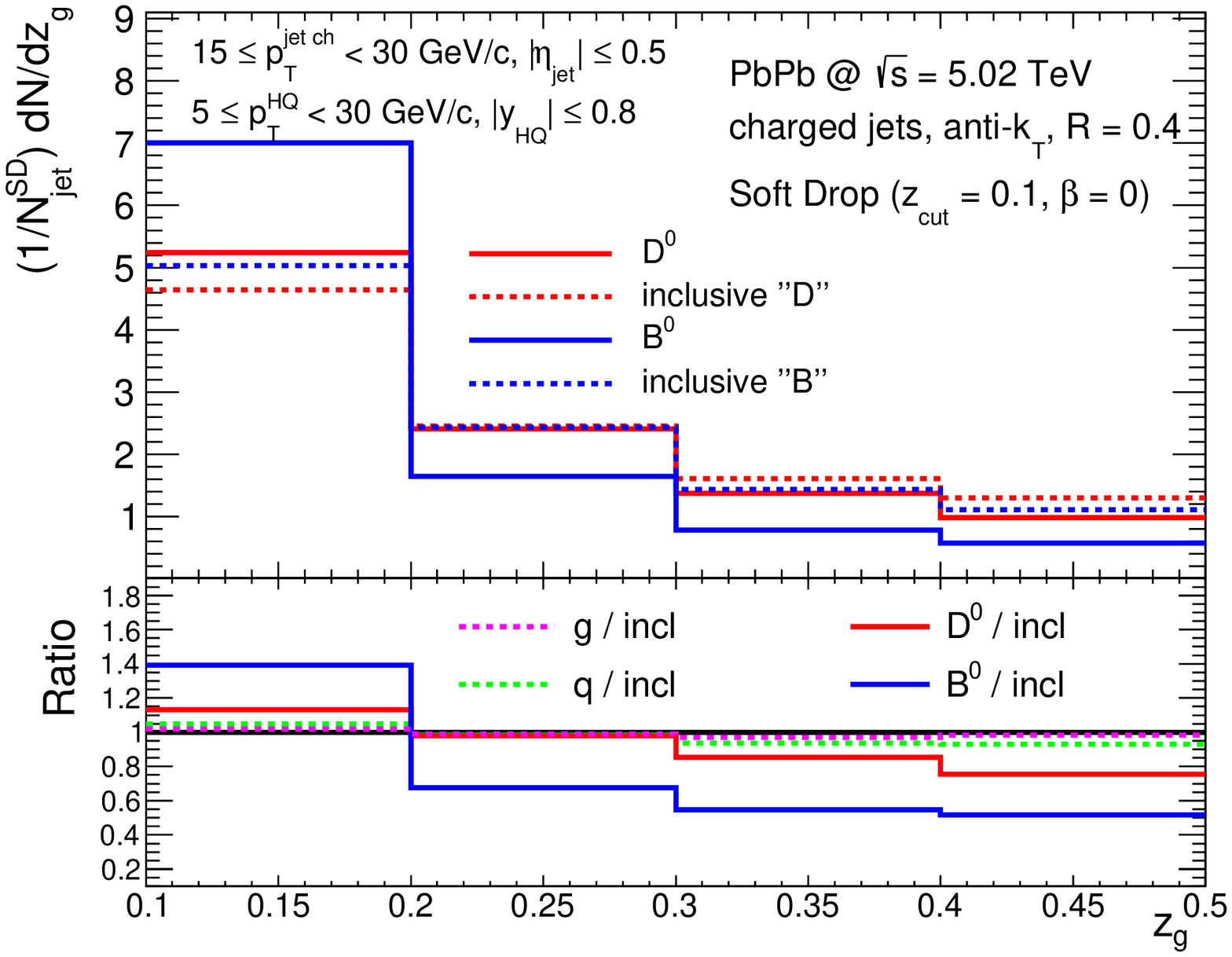}
				\includegraphics[width=0.49\textwidth]{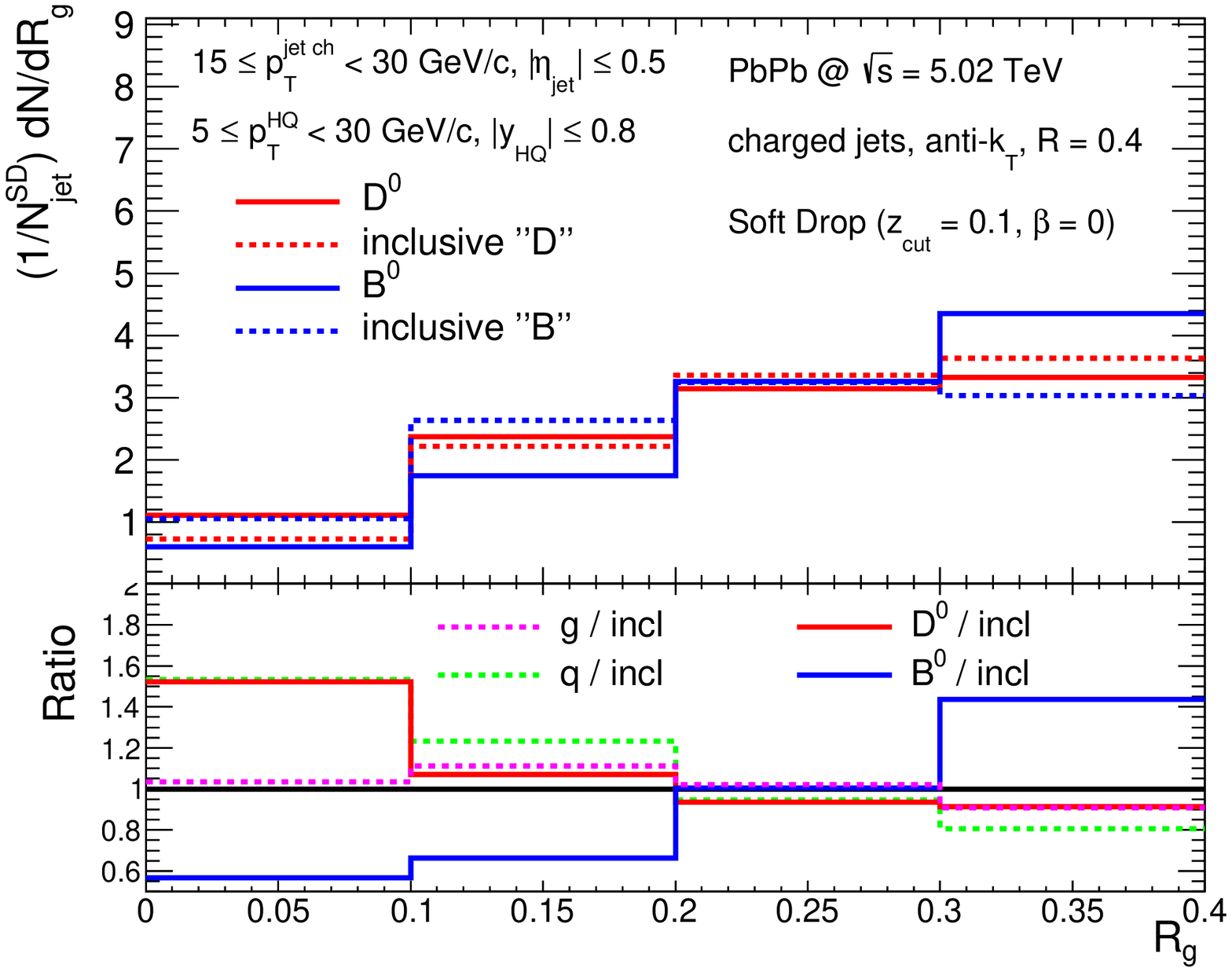}
		\end{center}
	\end{minipage}
	
	\caption{(color online) Upper: Predictions of $z_g$ (left) and $R_g$ (right) distributions for D$^0$-tagged, B$^0$-tagged jets and inclusive "D", inclusive "B" jets in 15 $\leq p_{\rm T}^{\rm jet\ ch} <$ 30 GeV/$c$ in central PbPb collisions at $\sqrt{s}$ = 5.02~TeV, normalised to the number of groomed jets $N_{\rm jet}^{\rm SD}$. Bottom: $z_g$ (left) and $R_g$ (right) distribution-ratios of pure light-quark initiated, pure gluon initiated, D$^0$-tagged and B$^0$-tagged jets over those of inclusive jets.}
	\label{fig:aa4}
\end{figure*}

\section{Jet substructure in heavy-ion collisions}
\label{sec:AA}

In $A$+$A$ collisions, both radiative and collisional energy loss mechanisms contribute to the total medium-induced parton energy loss. In order to describe the in-medium evolution of heavy and light partons simultaneously, the Simulating Heavy quark Energy Loss with Langevin equations (SHELL) model~\cite{Dai:2018mhw,Wang:2020ukj,Wang:2020bqz,Wang:2020qwe} is utilized in this work, where the propagation of heavy quarks are described by the modified Langevin equations~\cite{Cao:2013ita, Dai:2018mhw, Wang:2019xey, Wang:2020bqz, Wang:2020qwe},
\begin{eqnarray}
\label{eq:lang1}
&&\vec{x}(t+\Delta t)=\vec{x}(t)+\frac{\vec{p}(t)}{E}\Delta t \ ,\\
&&\vec{p}(t+\Delta t)=\vec{p}(t)-\Gamma(p)\vec{p} \Delta t+\vec{\xi}(t)-\vec{p}_g \, ,
\label{eq:lang2}
\end{eqnarray}
where $\Delta t$ is the time interval between each Monte Carlo simulation step, $\Gamma(p)$ is the drag coefficient and is related with the diffusion coefficient $\kappa$ by the fluctuation-dissipation relation~\cite{He:2013zua}: $\Gamma = \frac{\kappa}{2ET}=\frac{T}{D_s E}$. The spacial diffusion coefficient $D_s(2\pi T)=4$ is extracted through comparisons between theoretical calculations and experimental data~\cite{STAR:2014wif, Xie:2016iwq, CMS:2017qjw, ALICE:2018lyv} of final-state D-meson productions, which agree well with $D_s = \frac{4}{2\pi T}$ given by lattice QCD~\cite{Francis:2015daa,Brambilla:2020siz}. The stochastic term $\vec{\xi}(t)$ is white noise representing the random kicks obeying $\left \langle \xi^i(t)\xi^j(t') \right \rangle =\kappa \delta^{ij}\delta(t-t')$, and $\vec{p}_g$ is the momentum recoil due to the medium-induced gluon radiation which is implemented with the higher-twist approach~\cite{Guo:2000nz,Zhang:2003yn,Zhang:2003wk,Majumder:2009ge}:
\begin{eqnarray}
\frac{dN}{dxdk^{2}_{\perp}dt}=\frac{2\alpha_{s}C_sP(x)\hat{q}}{\pi k^{4}_{\perp}}\sin^2(\frac{t-t_i}{2\tau_f})(\frac{k^2_{\perp}}{k^2_{\perp}+x^2M^2})^4\ ,
\label{eq:dNdxdk2}
\end{eqnarray}
where $x$ and $k_\perp$ are the energy fraction and transverse momentum of a radiated gluon, $M$ is the mass of a parent parton, $\alpha_s$ is the strong coupling constant, $C_s$ is the quadratic Casimir in color representation, $P(x)$ is the vacuum splitting function~\cite{Wang:2009qb}, $\hat{q} \propto q_0(T/T_0)^3$ is the jet transport coefficient~\cite{Chen:2010te} and $q_0 = 1.5\ \rm{GeV^2/fm}$ is extracted from final-state hadron productions~\cite{Zhang:2022fau}, $\tau_f = 2Ex(1-x)/(k_{\perp}^2 + x^2M^2)$ is the gluon formation time. A lower cutoff $\omega_0 = \mu_D = \sqrt{4\pi\alpha_s}T$ of energy is set to avoid the spectra divergence at $x \rightarrow 0$. The two energy loss mechanisms, originated from gluon radiations and mutual collisions respectively, are constrained by two independent parameters: jet transport coefficient $\hat{q}_0$ and spacial diffusion coefficient $D_s$. Ideally they can be related by $\hat{q}=2\kappa$~\cite{Cao:2013ita}, which may not hold in a realistic QCD medium~\cite{Beraudo:2009pe, Prino:2016cni} so they are extracted separately in this work.

The initial parton distributions are provided by the MC Glauber Model~\cite{Miller:2007ri}. The space-time evolution of the QCD medium is provided by a (3+1)D viscous hydrodynamic model CLVisc~\cite{Pang:2012he,Pang:2013pma}. Since the hadron gas stage is not within our concern, we focus only on the energy loss as the jet propagates through the strongly coupled plasma defined at $T > T_c$ = 165 MeV.

\begin{figure*}[htbp]

	\begin{minipage}{1\linewidth}
		\begin{center}
				\includegraphics[width=0.49\textwidth]{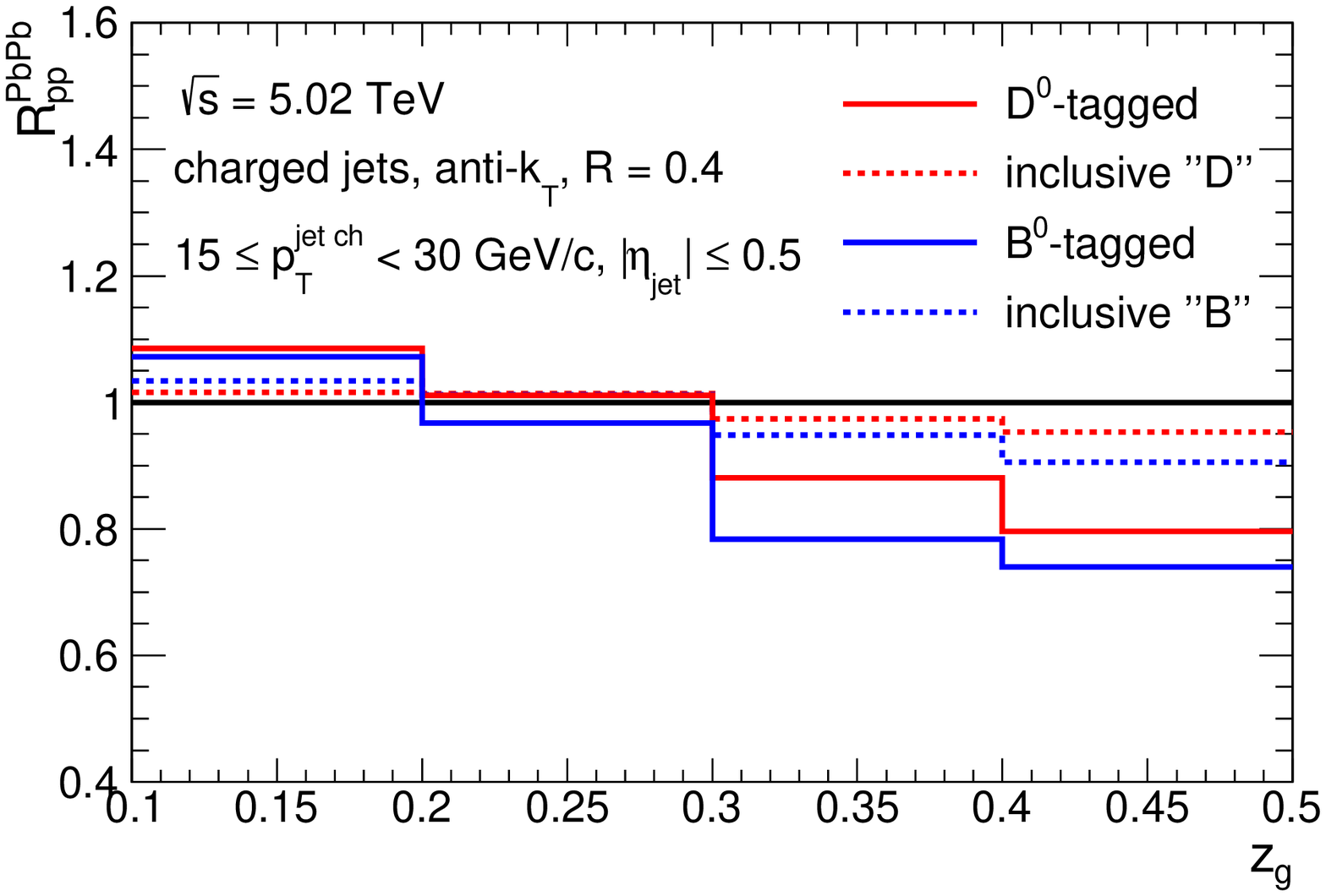}
				\includegraphics[width=0.49\textwidth]{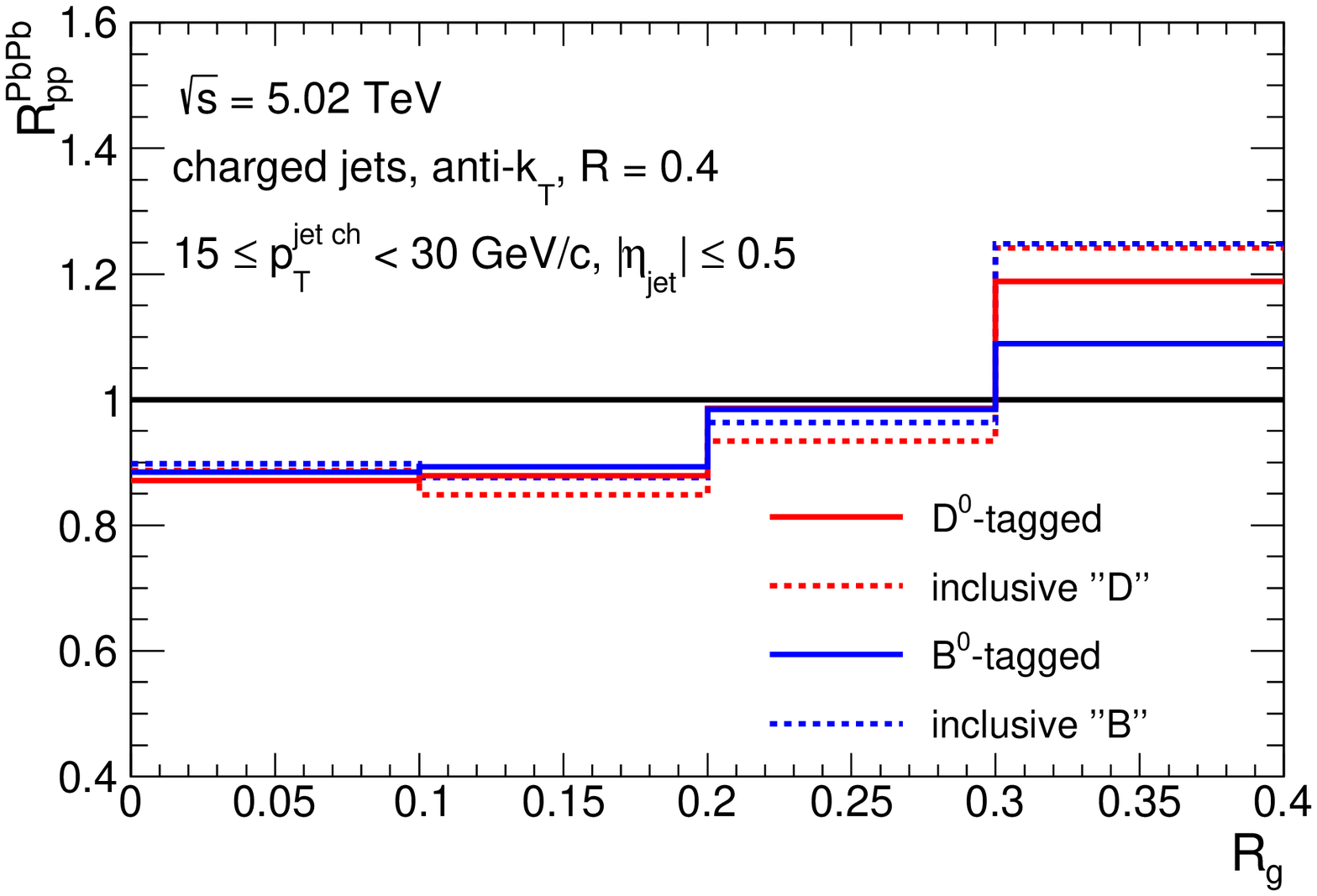}
		\end{center}
	\end{minipage}
	
	\caption{(color online) The PbPb/$pp$ distribution-ratios $R^{\rm PbPb}_{pp}$ of inclusive, D$^0$-tagged, B$^0$-tagged, light-quark initiated and gluon-initiated jets.}
	\label{fig:ratio4}
\end{figure*}

\begin{figure*}[htbp]

	\begin{minipage}{1\linewidth}
		\begin{center}
				\includegraphics[width=0.49\textwidth]{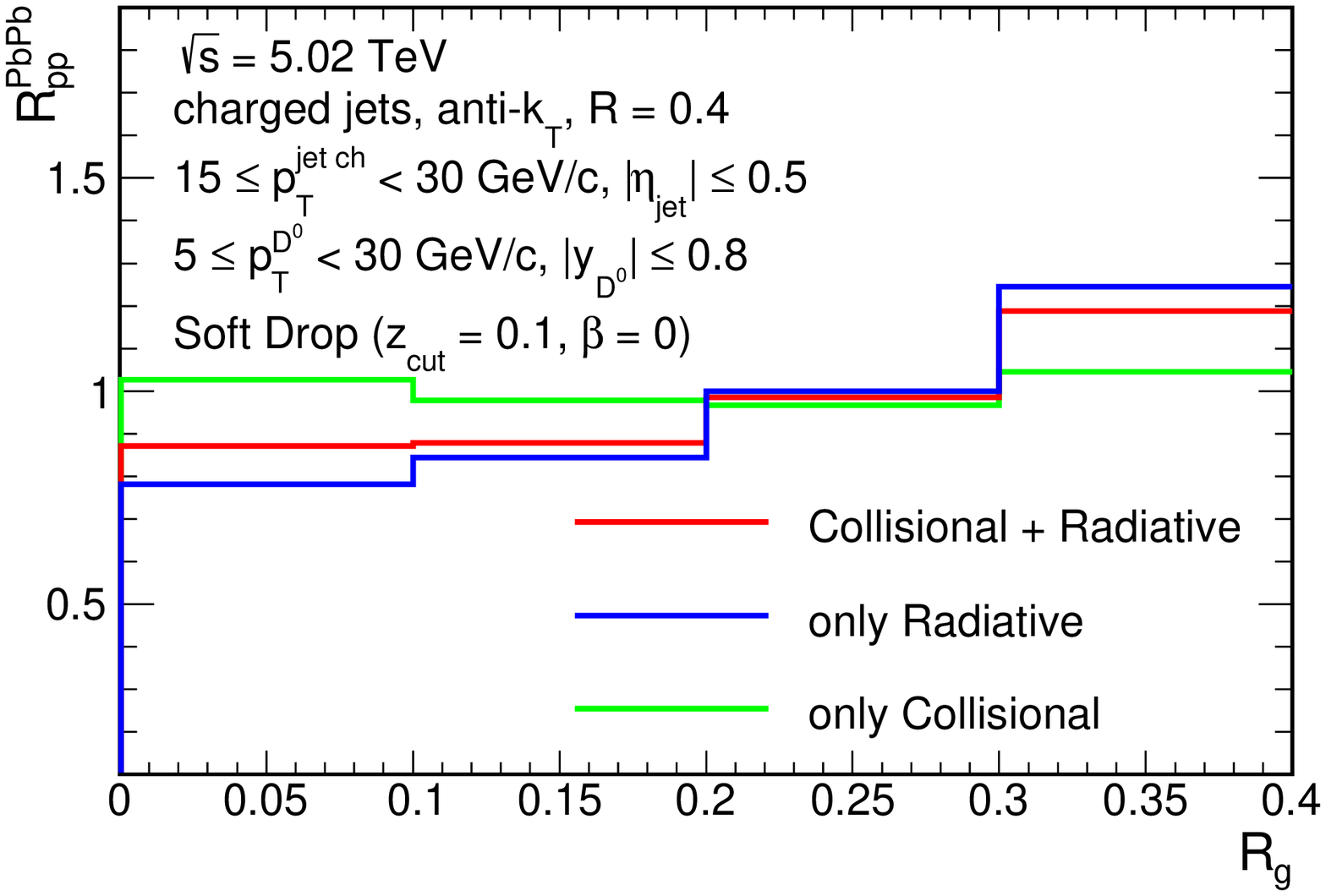}
				\includegraphics[width=0.49\textwidth]{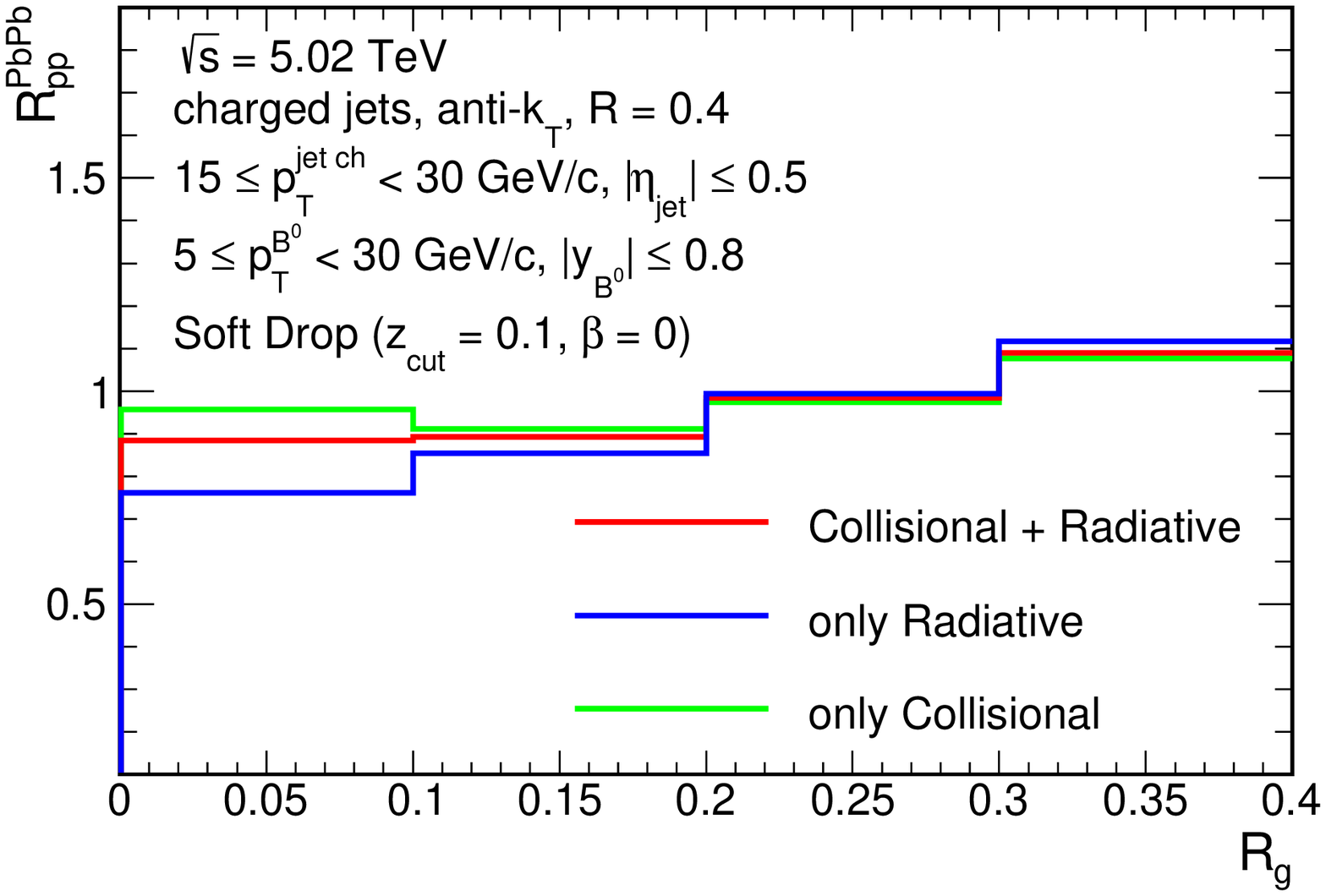}
		\end{center}
	\end{minipage}
	
	\caption{(color online) The PbPb/$pp$ distribution-ratios $R^{\rm PbPb}_{pp}$ of D$^0$-tagged (left) and B$^0$-tagged (right) jets concerning only collisional, only radiative, and collisional + radiative energy loss mechanisms.}
	\label{fig:frac}
\end{figure*}

Now we can predict the groomed jet substructure measurements in $A$+$A$ collisions. The upper panels of Fig.~\ref{fig:aa4} show the $z_g$ and $R_g$ distributions of B$^0$-tagged, D$^0$-tagged, inclusive jets "D" and inclusive jets "B" within the same kinematics range 15 $\leq p_{\rm T}^{\rm jet\ ch} <$ 30 GeV/$c$ in central PbPb collisions at $\sqrt{s}$ = 5.02~TeV, normalized to $N_{\rm jet}^{\rm SD}$. In the left panel, both B$^0$-tagged and D$^0$-tagged jets tend to distribute at smaller $z_g$ than their inclusive references, B$^0$-tagged jets distribute more than D$^0$-tagged jets at small $z_g$ and less at large $z_g$ the same as in $pp$ shown in Fig.~\ref{fig:pp4}. In order to further demonstrate the mass effects in these medium-involved jets produced in heavy-ion collisions, the ratios of B$^0$/incl, D$^0$/incl are also taken in the bottom panel where the mass hierarchy remained the same as in p+p. The $z_g$ ratios of q/incl and g/incl are still equal to unity within uncertainties as shown in the bottom panel confirming the fact that $z_g$ is not shifting much by the medium modifications. In the right panel, B$^0$-tagged jets distribute at larger angles than D$^0$-tagged jets originating solely from larger parton mass. $R_g$ distribution of B$^0$-tagged jets are wider than inclusive "B" while those of D$^0$-tagged jets are narrower than inclusive "D" which is the same case as in $pp$ collisions. We also look at the $R_g$ distributions of pure light-quark initiated and gluon-initiated jets demonstrated as ratios of q/incl and g/incl in the bottom panel. $R_g$ distributions for quark-initiated and gluon-initiated jets are more similar in $A$+$A$ collisions leading to the fact that $R_g$ for inclusive jets are almost identical to that for gluon-initiated jets. It indicates that in the measurements of splitting-angle distributions for jets, the Casimir colour factors are still competing with mass effects in the hot and dense medium. We also calculate the untagged rates for inclusive, D$^0$-tagged, and B$^0$-tagged jets, which are 4.4\%, 17.2\%, and 65.8\% respectively. They show strong mass dependence rather than medium modifications.

Next, we move on to the prediction of the medium modification factors of $z_g$ and $R_g$ distributions for inclusive, D$^0$-tagged, and B$^0$-tagged jets in central PbPb collisions at $\sqrt{s}$ = 5.02~TeV requiring 15 $\leq p_{\rm T}^{\rm jet\ ch} <$ 30 GeV/$c$, they are denoted as $R^{\rm PbPb}_{pp}$ which are plotted in Fig.~\ref{fig:ratio4}.
In the left panel, for all these kinds of jets, one can observe the distributions in $A$+$A$ are enhanced at small $z_g$ and suppressed at large $z_g$ compared to their corresponding $pp$ references indicating all the jets become more momentum-imbalanced due to the medium modifications. We find the larger the parton mass is, the more medium modifications these jets will suffer. In the right panel, all the $R_g$ distributions are suppressed at a small value and enhanced at a large value compared to the cases in $pp$, it shows that the splitting-angle distributions for these jets will become broadened due to the medium-induced modifications. We find the larger the parton mass is, the less medium modifications of splitting-angle these jets will suffer.

In our previous work we found that emission angle re-distributions in each splitting for D$^0$-tagged jets in $A$+$A$ collisions results from gluon emissions in QCD medium, namely as radiative energy loss mechanisms, and collisional energy loss mechanisms do not contaminate such distributions~\cite{Dai:2022sjk}. It's also very interesting to investigate the impacts of collisional energy loss mechanisms in the jet substructure observables since they are the hardest branching. 
The $R^{\rm PbPb}_{pp}$ of the normalized $R_g$ distributions for both D$^0$-tagged and B$^0$-tagged jets are plotted in Fig.~\ref{fig:frac} considering only collisional, only radiative and collisional + radiative energy loss mechanisms respectively. We can find the ratio only considering collisional energy loss equals unity indicating that collisional energy loss plays a negligible role in this angular redistribution of jet splittings and medium modifications to $R_g$ mainly come from radiative energy loss contributions.

\section{Summary}
\label{sec:sum}

The jet substructure observable, the momentum splitting fraction $z_g$ and the groomed jet radius $R_g$ of inclusive, D$^0$-tagged and B$^0$-tagged jets in both $pp$ and PbPb collisions at $\sqrt{s}=5.02$~TeV are predicted and investigated in the kinematics range 15 $\leq p_{\rm T}^{\rm jet\ ch} <$ 30 GeV/$c$. In $pp$ collisions, we find that the larger the quark mass is, the more momentum-imbalanced these jets will be. B$^0$-tagged jets distribute at larger $R_g$ than the D$ ^0$-tagged jets, however, D$ ^0$-tagged jets are narrower than their inclusive jets counterparts. The inclusive jets are mainly comprised of gluon-initiated jets, in this specific kinematics region the mass effects and the Casimir color effects are competing with each other since they both lead the jets to broaden. In the results of the predictions in $A$+$A$ collisions, all these jets become momentum imbalanced and splitting-angular wider than those in $pp$ due to jet quenching effects. The massive jets will suffer larger $z_g$ modifications and fewer $R_g$ modifications. When we study the mass effect in $A$+$A$ collisions, we find both the massive to the massless ratios are similar to that in $pp$, the mass hierarchy can still be kept in $z_g$ observable and the competition between mass effects and the Casimir color factor effect on $R_g$ distributions will also be found in those medium evolved jets in the same kinematics range. We further find that medium modifications to $R_g$ mainly come from radiative energy loss contributions instead of collisional energy loss.

~\\

{\bf ACKNOWLEDGEMENTS:} We thank Chi Ding and Xiang-Yu Wu for providing detailed profiles of the (3+1)D viscous hydrodynamic model CLVisc. This research is supported by the Guangdong Major Project of Basic and Applied Basic Research No. 2020B0301030008, the Natural Science Foundation of China with Project Nos. 11935007 and 11805167.

\vspace*{-.6cm}

\bibliographystyle{apsrev4-1}
\bibliography{ref}

\end{document}